\documentclass[aps,prd,twocolumn,amsmath,amssymb,showpacs,floatfix,nofootinbib,superscriptaddress]{revtex4-1}

\usepackage{graphicx}
\usepackage{dcolumn}
\usepackage{bm}
\usepackage{hyperref}
\usepackage{color}

\begin{document}

\title{Searching for Oscillations in the Primordial Power Spectrum with CMB and LSS Data}

\author{Chenxiao Zeng}
\email{zeng.544@osu.edu}
\affiliation{Department of Physics, The Ohio State University, Columbus, OH 43210, USA}
\affiliation{Department of Physics \& Astronomy, Johns Hopkins University, Baltimore, MD 21218, USA}
\affiliation{Center for Cosmology and Astroparticle Physics (CCAPP), The Ohio State University, Columbus, OH 43210, USA}
\author{Ely D. Kovetz}
\affiliation{Department of Physics \& Astronomy, Johns Hopkins University, Baltimore, MD 21218, USA}
\author{Xuelei Chen}
\affiliation{Key Laboratory of Computational Astrophysics, National Astronomical Observatories, Chinese Academy of Sciences, Beijing 100012, China}
\affiliation{School of Astronomy and Space Science, University of Chinese Academy of Sciences, Beijing 100049, China}
\affiliation{Center for High Energy Physics, Peking University, Beijing 100871, China}
\author{Yan Gong}
\affiliation{Key Laboratory of Computational Astrophysics, National Astronomical Observatories, Chinese Academy of Sciences, Beijing 100012, China}
\author{Julian B. Mu\~noz}
\affiliation{Department of Physics, Harvard University, Cambridge, MA 02138, USA}
\author{Marc Kamionkowski}
\affiliation{Department of Physics \& Astronomy, Johns Hopkins University, Baltimore, MD 21218, USA}

\begin{abstract}
Different inflationary models predict oscillatory features in the primordial power spectrum.
These can leave an imprint on both the cosmic microwave background (CMB) and the large-scale structure (LSS) of our Universe, that can be searched for with current data.
Inspired by the axion-monodromy model of inflation, we search for primordial oscillations that are logarithmic in wavenumber,  using both CMB data from the Planck satellite and LSS data from the WiggleZ galaxy survey. 
We find that, within our search range for the new oscillation parameters (amplitude, frequency and phase), both CMB-only and CMB+LSS data yield the same best-fit oscillation frequency of $\log_{10}\omega=1.5$,  improving the fit over $\Lambda$CDM by $\Delta\chi^2\! =\! -9$ and $\Delta\chi^2\! =\! -13$ (roughly corresponding to $2\sigma$ and $2.8\sigma$ significance), respectively. 
 Bayesian evidence for the log-oscillation model versus  $\Lambda$CDM  indicates a very slight preference for the latter. Future CMB and LSS data will further probe this scenario.
\end{abstract}

\maketitle

\section{\label{sec_Introduction}Introduction}

Cosmological inflation is the widely preferred paradigm to explain the origin of primordial fluctuations~\cite{Guth:1980zm, Linde:1981mu, Albrecht:1982wi}. In the simplest implementations, a single scalar field slowly rolling down a shallow potential (the inflaton) drives a rapid stage of exponential accelerated expansion in the early Universe, which imprints microscopic quantum fluctuations onto cosmic scales, seeding the growth of the large-scale structure. 

An appealing class of inflationary models naturally realized in string theory, known as axion monodromy~\cite{Silverstein:2008sg,McAllister:2008hb}, feature an inflaton field that is protected by a shift symmetry, which prevents large quantum corrections.
These models generically predict the presence of oscillations in the primordial power spectrum that are logarithmic in wavenumber~\cite{Flauger:2009ab,Behbahani:2011it}.  
Analogous effects were also discussed earlier in connection with Planck-scale or trans-Planckian effects \cite{Wang:2002hf,Martin:2000xs,Easther:2002xe,Danielsson:2002kx,Bozza:2003pr}, as well as unwinding inflation~\cite{DAmico:2012wal}.

The primordial fluctuations sourced during inflation induce curvature perturbations, which largely determine the statistical properties of the cosmic microwave background (CMB) and large-scale structure (LSS) of our Universe. 
Therefore, these tracers can provide useful test beds for distinguishing between possible models of inflation.
In particular, CMB data have been used to search for oscillatory features with WMAP~\cite{Pahud:2008ae,Meerburg:2013cla} and Planck power spectra~\cite{Planck:2013jfk,Meerburg:2013dla,Ade:2015coi}, as well as with measurements of the CMB bispectrum~\cite{Munchmeyer:2015ela,Fergusson:2015cps,Akrami:2018odb}.
Additionally, different authors have carried out numerical analysis with the past CMB and LSS data sets~\cite{Aich:2013oi, Hazra:2013} and laid out forecasts for constraints with future observations~\cite{Huang:2012mr,Chen:2016vvw,Ballardini:2016pp,Ballardini:2018pp,Escudero:2015yka,Palma:2018co}.

In this note, we present results from a search for log-oscillations in the primordial power spectrum using both CMB and LSS measurements, with the latest data from the Planck 2015 release~\cite{Ade:2015xua} and the WiggleZ galaxy survey~\cite{Parkinson:2012vd}.
We use a modified version of the standard Boltzmann code (\texttt{CLASS}{})~\cite{Lesgourgues:2011re} to calculate the observables in the presence of oscillations, and we constrain the oscillation parameters with a Monte Carlo Markov Chain (MCMC) analysis of the data, by implementing the Multimodal Nested Sampling Algorithm (\texttt{MultiNest})~\cite{Feroz:2007kg, Feroz:2008mn} via the \texttt{Monte Python} package~\cite{Audren:2012wb}.
Ours is the first search for log-space oscillations in the primordial power spectrum using nested sampling and joint CMB and LSS measurements. We find that the addition of three oscillation parameters (amplitude, frequency and phase) provides a significantly better fit to the data.  However, based on the Bayesian evidence, the log-spaced oscillations are not preferred over the bare $\Lambda$CDM model given current CMB+LSS data.

\section{\label{sec_physics_of_axion_monodromy}Oscillations from Axion Monodromy}

In single-field inflationary models, the primordial power spectrum
is directly related to the inflaton potential $V(\phi)$. 
We
consider an axion-monodromy potential given by~\cite{Flauger:2009ab}
\begin{equation}
    V(\phi) \equiv \mu^3\phi + \Lambda^4 \mathrm{cos}\Big( \frac{\phi}{f} \Big)
\end{equation}
where $\mu^3 \phi$ is the smooth potential in the axion model, $\Lambda^4$ is modulation amplitude, and $f$ the periodicity in $\phi$. 
The equation of motion that follows from this potential is then
\begin{equation}
\label{eq:phi}
\ddot{\phi}+3H\dot{\phi}+\mu^3- \dfrac{\Lambda^4}{f} \sin \left(\frac{\phi}{f}\right)=0\,.
\end{equation}
In the absence of oscillations, the primordial scalar power spectrum is given by $A_s(k/k_*)^{n_s-1}$, where  $A_s$ is the amplitude, $k_*$ is the pivot wavenumber, and $n_s$ the scalar spectral index.
Solving for the power spectrum using Eq.~\eqref{eq:phi} leads to the appearance of log-spaced oscillations,
\begin{equation}
\label{eq:Pk}
    P_{\mathcal{R}}(k) = A_s \left(\frac{k}{k_*}\right)^{n_s-1}  \left(1 + \epsilon \ \mathrm{cos} \left[\omega \ \mathrm{ln}\left(\frac{k}{k_*}\right)+\phi _k \right] \right),
\end{equation}
in the curvature power spectrum $P_{\mathcal{R}}(k)$, with oscillation frequency $\omega$, phase $\phi_k$, and perturbative amplitude $\epsilon$.
We will vary these three parameters in our analysis below.

\section{\label{sec_methodology}Methodology and Data Sets}

Given the primordial power spectrum from Eq.~\eqref{eq:Pk}, we can calculate the observed angular power spectrum of the CMB, and matter power spectrum of the LSS, through transfer functions.
We calculate these with the standard Boltzmann \texttt{CLASS}, and modify the primordial power spectrum according to Eq.~\eqref{eq:Pk} to obtain the CMB and LSS observables.
We sample the parameter space for the six $\Lambda$CDM parameters plus the three new oscillation parameters using \texttt{MultiNest} and the MCMC sampler \texttt{Monte Python}.
As we shall see, the likelihood of the oscillation frequency $\omega$ presents a large amount of local maxima, which hampers exploration of the phase space with standard MCMC samplers, as these tend to dwell in each maximum and not explore the whole parameter space. To circumvent this, we use \texttt{MultiNest} to establish constraints over the entire $\omega$ range, and roughly identify the global maximum-likelihood regions.
We then obtain the best-fit oscillation parameters and the corresponding $\chi^2$ values by running short standard Metropolis-Hastings chains with narrow priors on $\omega$ around the peaks of the \texttt{MultiNest} runs. 
We set flat priors on all the model parameters except $A_s$ and $\omega$, which have log-spaced priors, with ranges consistent with the analysis of Ref.~\cite{Ade:2015coi}, as shown in Table~\ref{table_parameters}\footnote{
We note that we had to augment the sampling precision of the primordial power spectrum in {\tt CLASS} by increasing the \texttt{k\_per\_decade\_primordial} parameter, which determines the number of points per decade in the primordial logarithmic $k$ space. 
We set this parameter to 5000, as the standard value does not resolve the oscillations of our model for large values of $\omega$.}.

\begin{table}[h]
\begin{center}
\caption{\label{table_parameters}
List of parameters with prior ranges for \texttt{MultiNest}. Before using multimodal nested sampling, we sample $\Lambda$CDM and the 14 foreground parameters using \texttt{Monte Python}. We then fix the foregrounds to their best-fit values and perform the \texttt{MultiNest} runs.
}
\begin{ruledtabular}
\begin{tabular}{lcdr}
\textrm{Model}&
\multicolumn{1}{c}{\textrm{Parameter}}&
\multicolumn{1}{c}{\textrm{Range [min, max]}}\\
\colrule
 & $100\Omega_b$ & [2.15, 2.30]\\
 & $\Omega_{cdm}$ & [0.115, 0.125]\\
$\Lambda$CDM & $100\theta_s$ & [1.03,1.05]\\
 & $\ln 10^{10}A_s$ & [3.00,3.18]\\
 & $n_s$ & [0.95,0.98]\\
 & $\tau_{reio}$ & [0.040,0.125]\\
\colrule
 & $\epsilon$ & [0.00,0.50]\\
Log-oscillation & $\log_{10}\omega$ & [0.00,2.10]\\
 & $\phi_k$ & [0.00,6.28]\\ 
\end{tabular}
\end{ruledtabular}
\end{center}
\end{table}

The data sets we use are the Planck 2015 TT\footnote{plik\_dx11dr2\_HM\_v18\_TT\_bin1.clik} and LowTEB\footnote{lowl\_SMW\_70\_dx11d\_2014\_10\_03\_v5c\_Ap.clik} likelihoods~\cite{Aghanim:2016cla}, as measurements of the CMB power spectrum, complemented by the WiggleZ final release data \cite{Parkinson:2012vd},
based on the $238,000$ galaxies observed with redshifts in the Southern sky between 2006 and 2011 by the 3.9-meter Anglo-Australian Telescope (AAT).
For the Planck data we maximize the sensitivity to very sharp features in the frequency by using the unbinned (``bin1") versions of the TT and lowTEB likelihoods instead of the standard binned versions.
We simultaneously vary the three oscillation parameters $\{\omega,\epsilon,\phi_k\}$ and  the six $\Lambda$CDM parameters $\{A_s,n_s,\omega_c,\omega_b,\theta_s,\tau\}$, while we fix the 14 foreground parameters to their best fits for $\Lambda$CDM, as in Ref.~\cite{Ade:2015xua}.
For WiggleZ we use the likelihood implemented in \texttt{Monte Python} for the low-redshift matter power spectrum, which should be sensitive to the oscillations that we are after.
Furthermore, the WiggleZ likelihood from Monte Python marginalizes over the galaxy bias analytically at each redshift (see e.g.\ Ref.~\cite{Escudero:2015yka}, Eq.~(11)).

To compare the base $\Lambda$CDM model with our three-parameter oscillatory extension, we compute the evidence and the Bayes factor for both cases. 
The Bayesian evidence can be expressed as
\begin{equation}
\label{eq_evidence}
    E(D|\mathcal{M}_i)  = \int d\theta_{ij}\mathcal{L}(D|\theta_{ij}, \mathcal{M}_i)\pi(\theta_{ij}|\mathcal{M}_i ),
\end{equation}
where $\theta_{ij}$ are the parameters defining the model $\mathcal{M}_i$, $\mathcal{L}(D|\theta_{ij}, \mathcal{M}_i)$ is the likelihood function for data $D$, and $\pi(\theta_{ij}|\mathcal{M}_i )$ is the prior of the parameters given a model.

The ratio of plausibilities of the two models $\mathcal{M}_a$ and $\mathcal{M}_b$ is used to define the Bayes factor between models $a$ and $b$ as
\begin{equation}
\label{eq_bayes_factor}
    \frac{P(\mathcal{M}_a|D)}{P(\mathcal{M}_b|D)} = \frac{E(D|\mathcal{M}_a)\pi(\mathcal{M}_a) }{E(D|\mathcal{M}_b)\pi(\mathcal{M}_b)  } \equiv B_{ab}.
\end{equation}
In practice, we compute the evidence Eq.~\eqref{eq_evidence} and Bayes factor Eq.~\eqref{eq_bayes_factor} by implementing \texttt{MultiNest} and calculating the posterior probabilities of the parameters of the two models ($\Lambda$CDM and the oscillatory extension), and we use the Jefferys' scale~\cite{Jeffreys:1939to} to interpret the Bayes factor as strength of evidence.

\section{\label{sec_results}Results}

\begin{figure}[b!]
\includegraphics[width=0.48\textwidth]{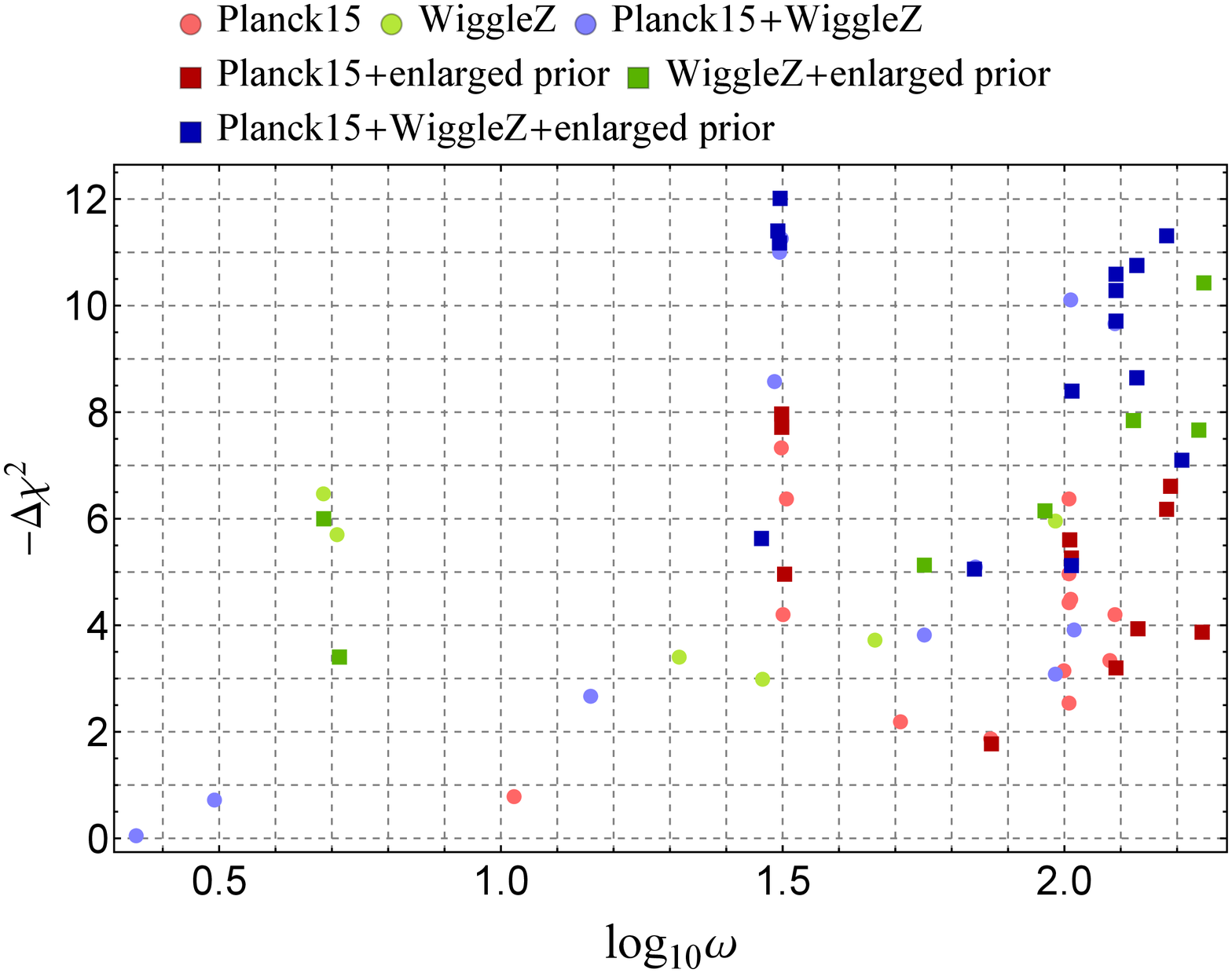}
 \caption{\label{fig_omega2}The frequency modes constrained by ``unbinned" Planck 2015 TT+lowTEB, WiggleZ~\cite{Parkinson:2012vd}, and their combination using \texttt{MultiNest}. Each dot gives the $\Delta\chi^2$ value at each peak in the parameter space of log frequency. We simultaneously constrain all nine parameters while fixing the foregrounds for each data set. Further validation is carried out by setting a flat instead of a log-spaced prior on the frequency with range $\omega=[0, 125.89]$ (darker squares) so as to potentially remove any nested-sampling dependence on the prior. With the complementary LSS data (blue dots), $\log_{10}\omega\approx 1.5$, the frequency suggested by the Planck paper~\cite{Ade:2015coi}, remains the most significant.}
\end{figure}

\begin{figure}
\includegraphics[width=0.48\textwidth]{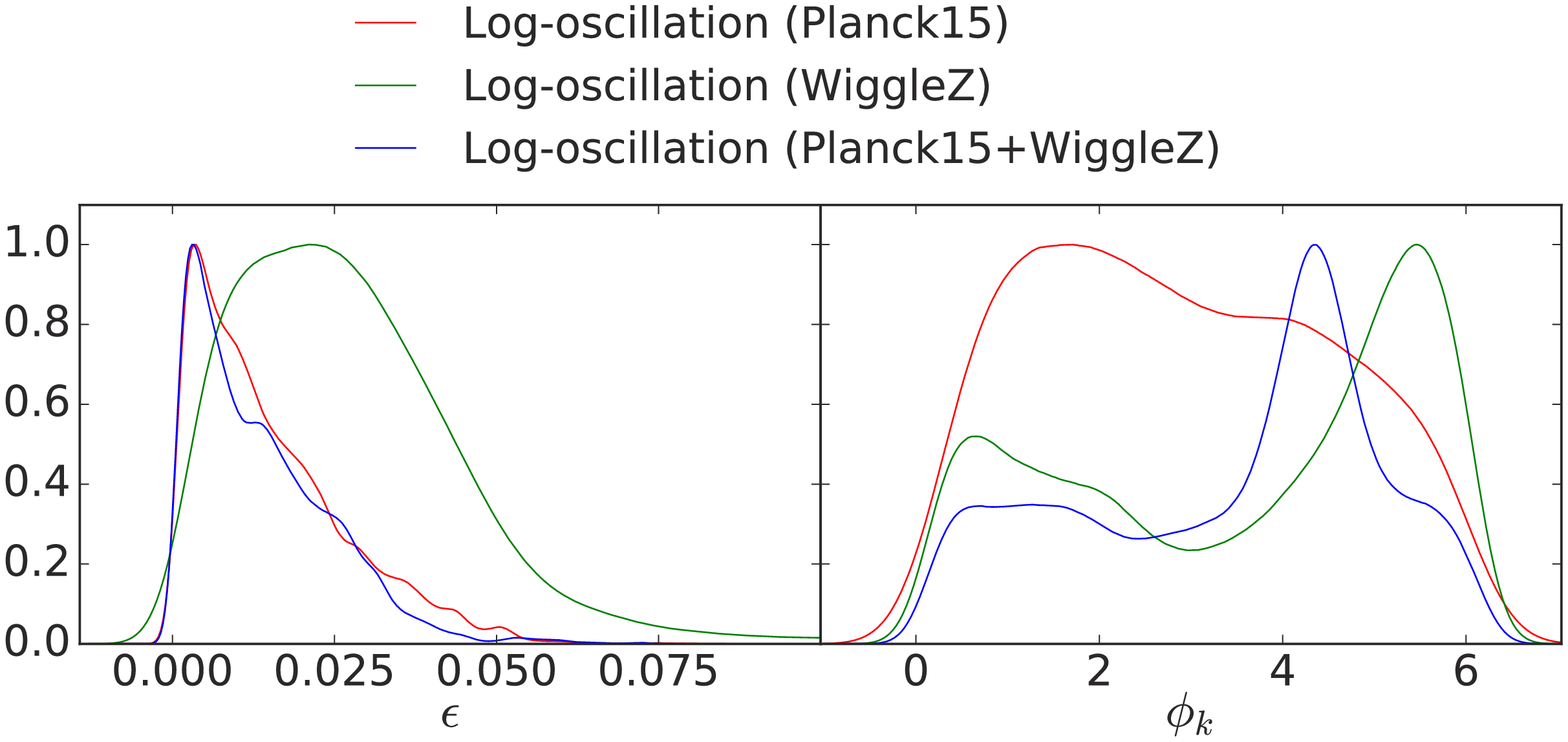}
\caption{\label{fig_omega} 1D posteriors for oscillation amplitude $\epsilon$ and phase $\phi_k$ with different data sets. We simultaneously constrain all nine parameters while fixing the foregrounds for each data set.}
\end{figure}

We show in Fig.~\ref{fig_omega2} the frequency modes identified by \texttt{MultiNest} and in Fig.~\ref{fig_omega} the posteriors for the other two oscillation parameters (amplitude $\epsilon$ and phase $\phi_k$). 
The likelihood for $\log_{10}\omega$
is expected to be highly multimodal, since the noise is comparable to the amplitude of the log-oscillatory feature.
This is indeed what we find in Fig.~\ref{fig_omega2}, where every dot represents the improvement of $\chi^2$ at a different frequency for the most likely modes. In this paper, $\Delta\chi^2$ refers to the $\chi^2$ of the log-oscillation model subtract that of $\Lambda$CDM.
We find that the {\tt MultiNest} algorithm finds modes with different improvement of $\chi^2$ at approximately the same frequencies if these are within a high-likelihood region. 
Using CMB data alone we find the most significant mode at $\log_{10}\omega \approx 1.5$, which is consistent with the Planck result of Ref.~\cite{Ade:2015coi}. 
Using WiggleZ data alone several modes appear significant, especially in the high-frequency region.
Interestingly, combining CMB and LSS power spectra results in the same high-likelihood peak at $\log_{10}\omega \approx 1.5$ with an increased likelihood over CMB-only data.
Additionally, the modes with $\log_{10}\omega \approx 2.0$ and $2.1$ also become pronounced. 
To ensure that our results are not extremely sensitive to our priors we ran the same chains with a flat prior $\omega = [0, 125.89]$, which we also show in Fig.~\ref{fig_omega2}.
We find that the frequency at $\log_{10}\omega \approx 1.5$ remains the most likely mode, with comparable $\Delta\chi^2$.

\begin{figure}[b!]
\includegraphics[width=0.49\textwidth]{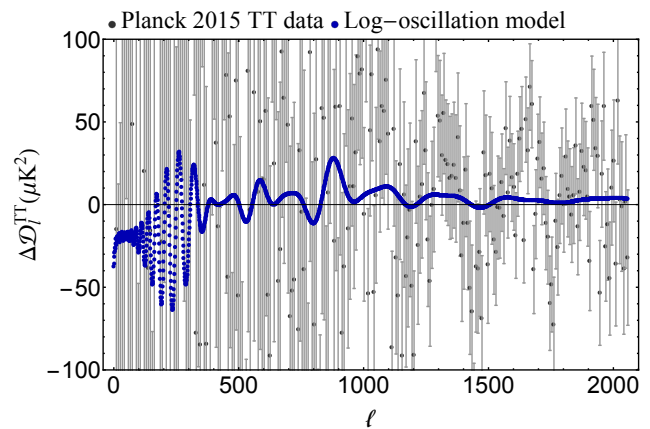}
\caption{\label{fig_cmbcl}The temperature power spectrum residuals of Planck 2015 \cite{Aghanim:2016cla} and the log-oscillation model (last row of Table~\ref{table_chi}), each with respect to the $\Lambda$CDM best fits (Table~\ref{table_chi}, third row).}
\end{figure}

\vspace{0.2in}
In Fig.~\ref{fig_cmbcl}, we show the CMB TT power spectrum residuals of Planck 2015, when compared to the best-fit $\Lambda$CDM model (third row of Table~\ref{table_chi}), as well as the change to said model when adding parameters given by the last row of Table~\ref{table_chi}.
The featured model induces oscillatory modulations in the CMB spectrum with largest offset around the first acoustic peak. These modulations are then greatly suppressed on small scales, although remain visible up to $\ell=2000$.

\begin{figure}
\includegraphics[width=0.46\textwidth]{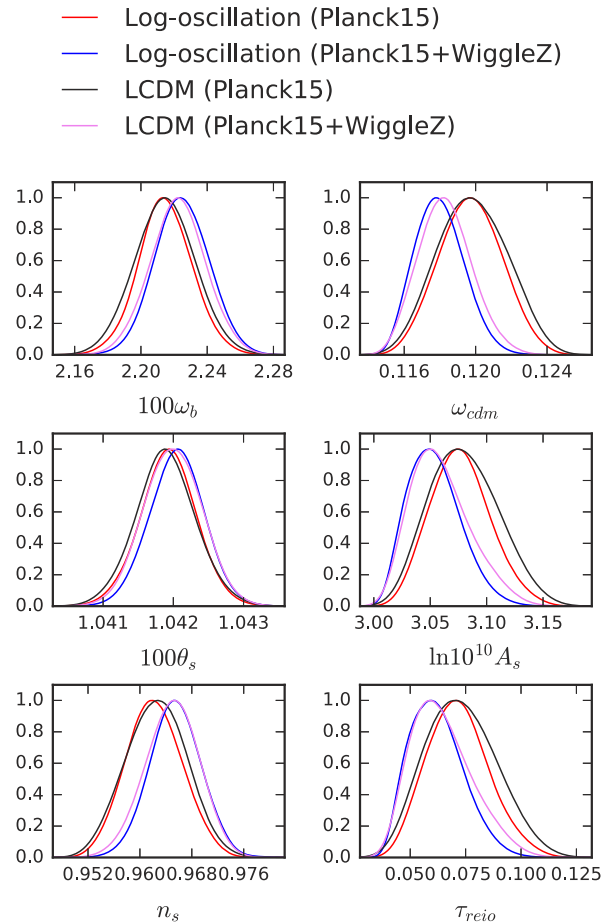}
\caption{\label{fig_posteriors}Comparison of the six parameters in $\Lambda$CDM (blue and green) and log-oscillation (black and red) posteriors with broad prior range of $\omega$. Legends on the top show the different model-data set combinations. The foreground parameters are fixed to the best-fits in the \texttt{Monte Python} results, and we vary the nine parameters simultaneously.}
\end{figure}

\begin{figure*}
\centering
\includegraphics[width=0.99\textwidth]{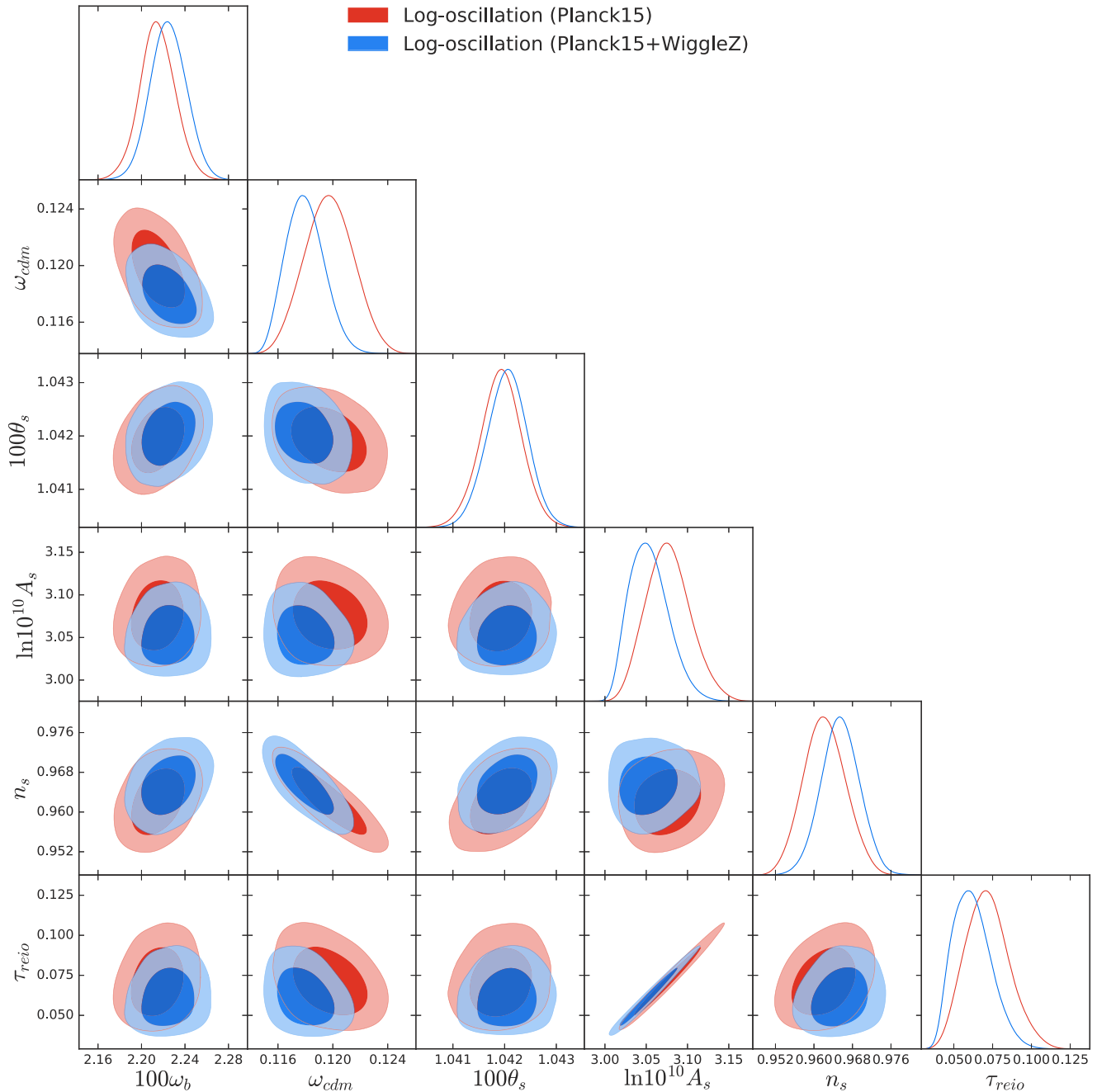}
\caption{\label{fig_triangle}2D posteriors for the $\Lambda$CDM parameters under the oscillation-based model, jointly constrained by CMB and LSS data using \texttt{MultiNest}. We vary the nine parameters simultaneously.}
\end{figure*}

Based on the peaks on $\Delta \chi^2$ from the \texttt{MultiNest} samplings on Fig.~\ref{fig_omega2}, we run short Metropolis-Hastings chains around each mode to estimate the best-fit parameter values for each model-data set combination, which are shown in Table~\ref{table_chi}.
\begin{table*}
\caption{\label{table_chi}The Bayes factor, improvements of $\chi^2$, and best-fits for different model-data set combinations. The likelihoods in the second column are given at the best-fit values, determined with an MCMC run around the peaks in the \texttt{MultiNest} results. $M_0$ and $M_1$ represent $\Lambda$CDM and the featured model, respectively. We confine the prior range of $\epsilon$ to $[0, 0.075]$ when calculating the Bayes factor. A small negative Bayes factor indicates a very slight preference (nearly no preference) for $\Lambda$CDM.}
\begin{ruledtabular}
\begin{tabular}{ccccccccccc}
 Models&$\Delta\log{E},\Delta\chi^2$&$100\omega_b$&$\omega_{cdm}$
&$100\theta_s$&$\ln10^{10}A_s$&$n_s$&$\tau_{reio}$&$\epsilon$&$\log_{10}\omega$&$\phi_k/2\pi$\\ \hline
 $M_0$(CMB)& &$2.21$ &$0.120$&$1.0421$&$3.07$&$0.963$&$0.069$& & & \\
 $M_1$(CMB)&$-0.312\pm0.180,\hspace{4.5pt} -8.67$&$2.23$&$0.118$&$1.0421$&$3.09$&$0.973$&$0.078$&$0.0248$&$1.50$&$0.651$\\
 $M_0$(CMB+LSS)& &$2.22$ &$0.118$&$1.0420$&$3.05$&$0.966$&$0.061$& & & \\
 $M_1$(CMB+LSS)&$-0.060\pm0.186, -12.68$&$2.24$&$0.118$&$1.0422$&$3.04$&$0.974$&$0.056$&$0.0291$&$1.50$&$0.698$\\
\end{tabular}
\end{ruledtabular}
\end{table*}
We find that including the three extra oscillation parameters in Eq.~\eqref{eq:Pk}, with $\log_{10}\omega\approx 1.5$ yields fit improvements corresponding to an effective $\Delta \chi^2=-8.67$, and $-12.68$ for the best fits of two data sets (CMB and CMB+LSS, respectively). The additional LSS data yield the best fit for nearly the same frequency as for the CMB data alone, showing that indeed the combination of data sets reinforces the presence of this oscillatory mode, corresponding to $2.8\sigma$ significance.  
Out of the three oscillation parameters, the major contribution to $\Delta \chi^2$ comes from the frequency $\omega$ and the amplitude $\epsilon$, while the effect of $\phi_k$ is negligible. The remaining $\Lambda$CDM parameters are not dramatically changed, as demonstrated in Figs.~\ref{fig_posteriors} and \ref{fig_triangle}, which show the one- and two-dimensional posteriors  with different model-data set combinations, respectively.
As expected, WiggleZ only marginally improves the constraints on the six $\Lambda$CDM parameters.

Although the fit of both CMB-only and CMB+LSS data is improved in the featured model, this does not necessarily represent a statistically significant result. 
We have to account for the larger volume of the oscillatory parameter space when computing the evidence $E$.
For our choice of priors in Table~\ref{table_parameters}, the corresponding Bayes factors between $\Lambda$CDM and the featured model are $B=-2.061\pm0.186$ (CMB) and $B=-1.655\pm0.192$ (CMB+LSS). 
The Bayes factor for the CMB-only constraint is consistent with the Planck Collaboration's result $-1.9$ \cite{Ade:2015coi}, where the negative value indicates that the standard $\Lambda$CDM scale-invariant power spectrum is preferred over log-oscillation. 
We note, though, that from Fig.~\ref{fig_omega} the posterior of $\epsilon$ only spans a small range of the default prior $[0, 0.5]$, so the Bayes factor may be overly suppressed by our prior choice.
As a test, we confine the prior range to the $[0, 0.075]$ range, which is more representative of the oscillations preferred by the data,
and obtain Bayes factors of $B=-0.312\pm0.180$ (CMB) and $B=-0.060\pm0.186$ (CMB+LSS) as shown in Table~\ref{table_chi}. 
Even when decreasing the prior on $\epsilon$, we find a slight preference for $\Lambda$CDM over log-oscillation, both with CMB-only and CMB+LSS data, albeit much less pronounced than in the wide-prior case.

\section{\label{sec_discussion_conclusions}Conclusions}

In this work, we used recent CMB and LSS data releases to constrain oscillation parameters in the primordial power spectrum.  
The main motivation to modify the form of the primordial power spectrum relies on the fact that small modulation to the nearly scale-invariant form may suggest the presence of an underlying symmetry obeyed by the inflaton field. Specifically, we considered an alternative form of the  power spectrum which is modulated by log-spaced oscillation features, generically sourced by models such as axion-monodromy inflation.

We have found, using CMB-only data, a preference for oscillations with a frequency $\log_{10}\omega\approx1.50$, at the $\Delta\chi^2 = -9$ level, in agreement with previous studies. Interestingly, the addition of LSS data from the WiggleZ survey reinforces that frequency as the best-fit mode and improves the fit by $\Delta\chi^2 = -13$, corresponding to $2.8\sigma$ significance.
Nonetheless, comparing the Bayesian evidence for the featured and standard $\Lambda$CDM models shows that current data have no strong preference for either, given the large amount of freedom in the three new oscillatory parameters.
In the future, it will be promising to revisit this analysis by using a combination of data from upcoming experiments such as CMB-S4~\cite{Abazajian:2016lcsb} or the Simons Observatory \cite{Ade:2018sbj} on the CMB side, and the Large Synoptic Survey Telescope (LSST)~\cite{Abell:2009lsb}, Dark Energy Spectroscopic Instrument (DESI)~\cite{Aghamousa:2016tde}, and Euclid~\cite{Laureijs} on the LSS front.

\acknowledgements
We thank Mario Ballardini, Liang Dai, Cora Dvorkin, Deanna Hooper, Tanvi Karwal, Georges Obied, Vivian Poulin, Sunny Vagnozzi and Hao-Yi Wu for very helpful discussions. This work was supported by NSF grant no.\ 0244990, NASA  NNX17AK38G, NSFC Grant No. 11633004, No. NSFC-11773031, No. NSFC-11822305, CAS Grant No. QYZDJ-SSW-SLH017, as well as the Simons Foundation. C.~Z. thanks Cora Dvorkin and the Harvard Physics Department for hospitality during its completion.

\end{document}